\documentclass[9pt,twocolumn,twoside]{pnas-new}
\usepackage{xcolor}
\usepackage{units}

\templatetype{pnasresearcharticle} 

\title{Emergence of topological electronic phases in elemental lithium under pressure}

\author[a,c]{Stephanie A. Mack}
\author[b,c]{Sin\'{e}ad M. Griffin} 
\author[a,b,c,d,1]{Jeffrey B. Neaton}

\affil[a]{Department of Physics, University of California Berkeley, Berkeley, CA 94720, USA}
\affil[b]{Molecular Foundry, Lawrence Berkeley National Laboratory, Berkeley, CA 94720, USA}
\affil[c]{Materials Sciences Division, Lawrence Berkeley National Laboratory, Berkeley, CA 94720, USA}
\affil[d]{Kavli Energy Nanosciences Institute at Berkeley, Berkeley, CA 94720, USA}

\leadauthor{Mack}

\authorcontributions{S.A.M., S.M.G., and J.B.N. designed research, performed research, analyzed data, and wrote the paper.}
\authordeclaration{The authors declare no conflict of interest.}
\correspondingauthor{\textsuperscript{1}To whom correspondence should be addressed. E-mail: jbneaton${@}$berkeley.edu}


\begin{abstract}
Lithium, a prototypical simple metal under ambient conditions, has a surprisingly rich phase diagram under pressure, taking up several structures with reduced symmetry, low coordination numbers, and even semiconducting character with increasing density. Using first-principles calculations, we demonstrate that some predicted high-pressure phases of elemental Li also host topological electronic structures. Beginning at 80 GPa and coincident with a transition to the \textit{Pbca} phase, we find Li to be a Dirac nodal line semimetal. We further calculate that Li retains linearly-dispersive energy bands in subsequent predicted higher pressure phases, and that it exhibits a Lifshitz transition between two \textit{Cmca} phases at 220 GPa. The \textit{Fd$\bar{3}$m} phase at 500 GPa forms buckled honeycomb layers that give rise to a Dirac crossing 1 eV below the Fermi energy. The well-isolated topological nodes near the Fermi level in these phases result from increasing p-orbital character with density at the Fermi level, itself a consequence of rising 1s core wavefunction overlap, and a preference for nonsymmorphic symmetries in the crystal structures favored at these pressures. Our results provide evidence that under pressure, bulk 3D materials with light elements, or even pure elemental systems, can undergo topological phase transitions hosting nontrivial topological properties near the Fermi level with measurable consequences; and that, through pressure, we can access these novel phases in elemental lithium.
\end{abstract}

\dates{Submitted on December 17, 2018; accepted by Proc. Natl. Acad. Sci. on March 26, 2019.}

\begin{document}

\verticaladjustment{-2pt}

\maketitle
\ifthenelse{\boolean{shortarticle}}{\ifthenelse{\boolean{singlecolumn}}{\abscontentformatted}{\abscontent}}{}

There has been considerable recent interest in understanding the phases of crystalline materials in terms of their topology as well as symmetry. This has led to the classification of new topological ground states, including topological insulators \cite{Hasan/Kane:2010}, Dirac/Weyl semimetals  \cite{Wan_et_al:2011,Young_et_al:2012,Soluyanov:2015}, and Dirac/Weyl nodal semimetals \cite{Kim_et_al:2015}, signatures of which have subsequently been observed in experiments. Structural phases with topological electronic structure have a host of interesting properties, including high mobilities, giant magnetoresistance, and chiral anomalies. Thus far, the majority of materials identified or verified experimentally that have topological electronic structures are binary or ternary compounds that include heavy elements \cite{Weng_et_al:2016}. Some elemental systems have also been predicted to exhibit topological nodal line properties at standard or high pressures, including Ca, Sr, Y \cite{Hirayama_et_al:2017}, Be, Mg \cite{Li_et_al:2016b}, and even the lightest elemental solid, H \cite{Naumov_et_al:2013,Naumov/Hemley:2016}. Yet, the majority of these elemental solids do not exhibit topological features at the Fermi energy and, if they do, these features are impeded by trivial bands. 

Lithium assumes a close-packed structure under ambient conditions and is a simple metal with a nearly-free electron-like band structure. Almost two decades ago, it was predicted using first-principles calculations that, somewhat counter-intuitively, as the pressure increases and the average electron density rises, lithium undergoes a sequence of phase transitions in which the coordination number decreases and the electronic structure strongly departs from that of a simple metal, culminating in a zero-gap semiconducting electronic structure around 100 GPa \cite{Neaton/Ashcroft:1999}. This series of phase transitions was attributed to a Peierls-like set of symmetry-lowering distortions \cite{Neaton/Ashcroft:1999}. At high but experimentally achievable densities, increasing overlap between 1s core states forces 2s valence electron density to reside in the interstitial regions \cite{Neaton_et_al:2001}, leading to an increase in p-orbital character of the band structure \cite{Boettger/Trickey:1985} near the Fermi energy \cite{Rousseau/Ashcroft:2008}. The kinetic energy of the increasingly nonuniform valence charge density is subsequently lowered via transitions to open structures with reduced coordination numbers. Interestingly, the initially-predicted zero-gap semiconducting phases exhibited a nonsymmorphic space group, \textit{Cmca}, with linearly dispersive bands \cite{Neaton/Ashcroft:1999} indicative of massless Dirac fermionic behavior with band velocities comparable to graphene \cite{Wallace:1947,Neto_et_al:2009}.

Since the original predictions nearly two decades ago \cite{Neaton/Ashcroft:1999}, experiments have confirmed that lithium exhibits lower coordinated structures with pressure in a diverse range of measured and predicted phases. At around 40 GPa, lithium has been measured to transform from close-packed \textit{Fm$\bar{3}$m} to a lower symmetry \textit{I$\bar{4}$3d} \cite{Hanfland_et_al:2000} phase in which the atoms are only three-fold coordinated. Subsequent \textit{ab initio} calculations \cite{Lv_et_al:2011,Pickard/Needs:2009} and experiments \cite{Guillaume_et_al:2011} have shown this three-fold coordination persists in a sequence of different structural phases, with nonsymmorphic symmetries, up to pressures of at least 450 GPa after which Li becomes four-fold coordinated. The sequence of low symmetry space groups Li is predicted to adopt at high pressure are mostly nonsymmorphic, which promote band sticking, increasing the likelihood of nontrivial topology and hosting novel band structure features, in spite of the fact that Li is the third-lightest element in the periodic table with negligible spin-orbit coupling.

Here, we use first-principles density functional theory (DFT) calculations to examine the most current and accepted predicted structures in the pressure phase diagram of Li. We compute and analyze their electronic structure and demonstrate that, in the proposed high-pressure structural phases of elemental Li, topologically-nontrivial electronic structures emerge under pressure. We calculate a change in the topological properties of the near-Fermi level band structure coinciding with the predicted structural phase transition at 80 GPa from the trivially insulating \textit{Aba2} phase to the \textit{Pbca} phase which forms a Dirac nodal line semimetal. The nodal ring in the \textit{Pbca} phase at the Fermi energy is well-separated from trivial bands by 0.8 eV in our DFT calculations. We further show that the structural transition between two predicted higher pressure \textit{Cmca} phases is also accompanied by a Lifshitz transition, wherein the Fermi surface topology changes from a single Dirac nodal ring at the Fermi level in the \textit{Cmca-24} phase to two nodal rings in the higher pressure \textit{Cmca-56}. The higher pressure \textit{P4$_2$/mbc} phase is predicted to have a nodal ring just below the Fermi energy. Lastly we compute that the highest pressure predicted phase \textit{Fd$\bar{3}$m} features a distorted hexagonal honeycomb network in strong analogy with graphene and a Dirac crossing 1 eV below the Fermi energy.

\subsection*{Results}

\begin{figure}
\includegraphics[width=\linewidth]{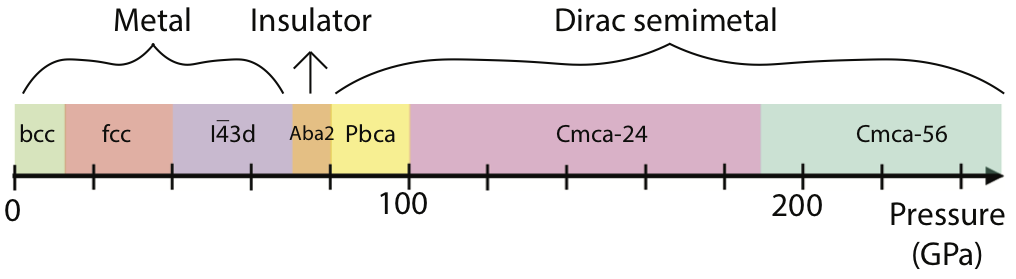}
\caption{\label{fig:pressure-struc} Predicted phase diagram of Li below 250 GPa from Ref.\cite{Lv_et_al:2011} at zero temperature. Note that from 100--165 GPa lithium assumes a \textit{Cmca} unit cell with 24 atoms per unit cell, and a 56 atom unit cell from 165--220 GPa, after which \textit{P4$_2$/mbc} is the preferred space group.}
\end{figure}

Experimental determination of the high-pressure structures of lithium is challenging for diffraction experiments due to its small atomic number (Z = 3). In fact, only four lower-pressure structural phases (\textit{Im$\bar{3}$m}, \textit{Fm$\bar{3}$m}, \textit{I$\bar{4}$3d}, \textit{Aba2}) have been experimentally confirmed. Beyond 70 GPa, although the Pearson class has been determined experimentally, the full crystallographic symmetry of the high pressure phases is not yet known \cite{Guillaume_et_al:2011}. Based on the Pearson class and using first-principles calculations, two prior studies \cite{Lv_et_al:2011,Pickard/Needs:2009} used structure searching algorithms to determine the low-enthalpy space group symmetries at a range of pressures and predicted the zero-temperature phase diagram of lithium from 0 to 500 GPa. We adopt structures from one of these studies \cite{Lv_et_al:2011}, which is in good agreement with an earlier independent study by Pickard and Needs \cite{Pickard/Needs:2009}; the difference between the two studies is that Ref. \cite{Lv_et_al:2011} predicts two intermediary structures at 71 and 227 GPa, not reported in \cite{Pickard/Needs:2009}. Starting with atomic coordinates from \cite{Lv_et_al:2011}, we use structures at a representative pressure in the predicted phase diagram (see SI), and then use DFT to compute and analyze their electronic structure. All DFT calculations are performed within the local density approximation (LDA) with PAW potentials treating all three electrons in Li as valence using the VASP code \cite{Kresse/Hafner:1993,Kresse/Furthmuller:1996}. We use the post-processing software Z2Pack \cite{Gresch_et_al:2017,Soluyanov/Vanderbilt:2011} and WannierTools \cite{Wu_et_al:2017} to compute topological invariants and surface states, and to determine whether a given phase is topologically trivial or nontrivial. Further details of these calculations can be found in the SI.

\begin{figure}
\includegraphics[width=\linewidth]{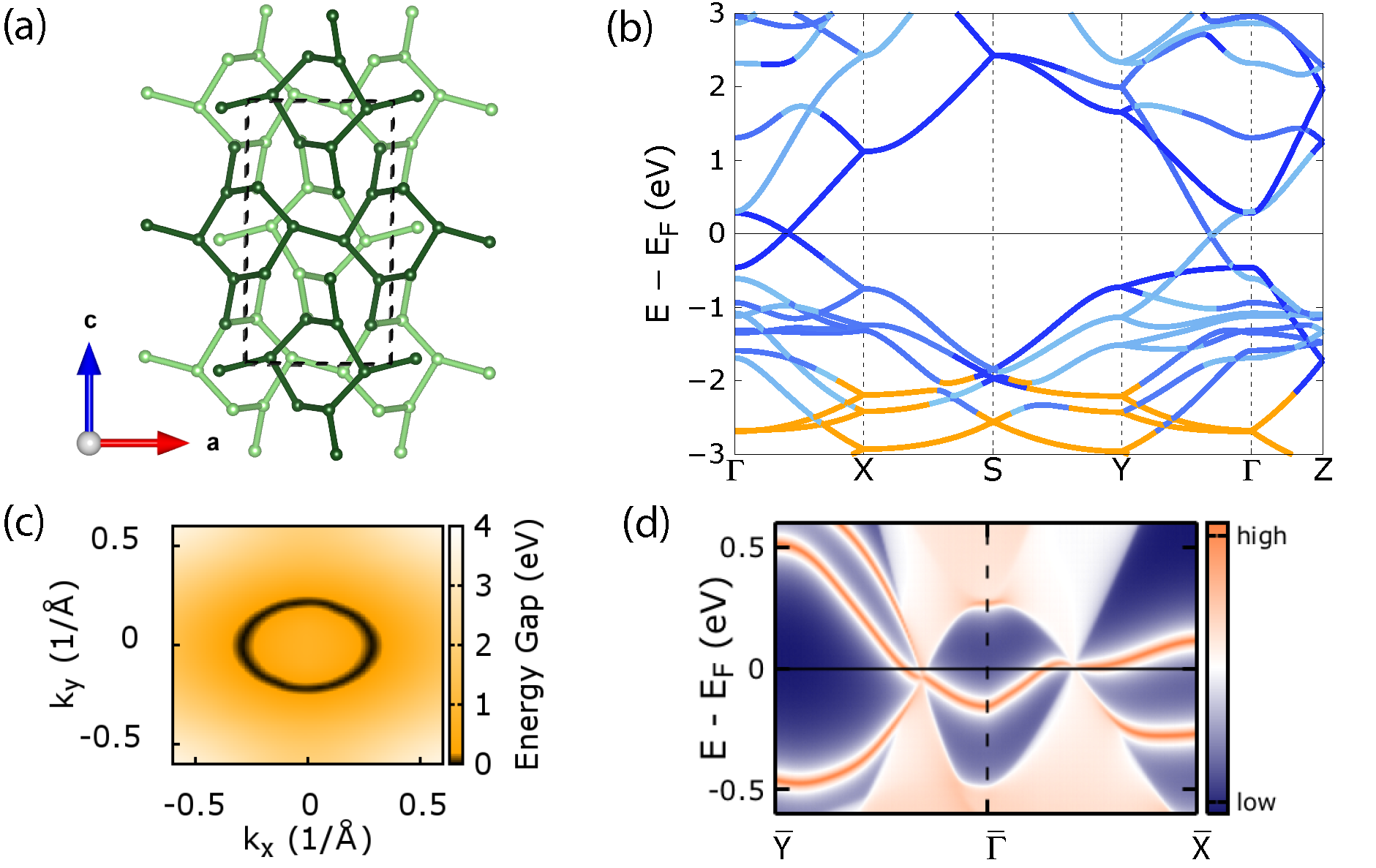}
\caption{\label{fig:pbca} Solid Li at 80 GPa in its predicted \textit{Pbca} phase. (a) \textit{Pbca} unit cell - coloring indicates different atomic layers along the $\boldmath{b}$ direction.  (b) Band structure at 80 GPa, where the colors indicate the dominant orbital contribution to the band, with the s, p$_x$, p$_y$, and p$_z$ orbitals represented by orange, light, medium, and dark blue respectively. The Fermi energy is set to 0 eV and marked by a black line. (c) The nodal ring is shown in the k$_z$ = 0 plane where the color gradient represents the size of the band gap in eV. The projection of the Fermi surface on the 2D k-plane indicates there is a nodal ring in the k$_z$=0 plane (d) The projected band structure along the [001] direction. See SI for details of our surface state calculations.}
\end{figure}

Although solid lithium is metallic \cite{Seitz:1935} and considered nearly-free-electron-like at low pressures, between 70-80 GPa it is predicted to adopt the \textit{Aba2} phase, which is predicted to be insulating (see FIG. \ref{fig:pressure-struc}). Using Z2Pack, we compute \textit{Aba2} to be topologically trivial with a $\mathbb{Z}_2$ index of zero. At 80 GPa, a transition to the semimetallic \textit{Pbca} phase is predicted; the crystal structure of \textit{Pbca} at 80 GPa is shown in FIG. \ref{fig:pbca}(a). Our calculated DFT band structure for \textit{Pbca} at 80 GPa (see SI for lattice parameters) features two four-fold degenerate Dirac points at the Fermi energy located along the $\Gamma$--X and $\Gamma$--Y directions as shown in FIG. \ref{fig:pbca}(b). Interestingly, these Dirac points are isolated from other bands. At the Dirac points (and away from the Fermi level at X and Y) we observe `band sticking' \cite{Young/Kane:2015}, or band degeneracies, enforced by the nonsymmorphic symmetries in the $Pbca$ space group. Similar band degeneracies arise in the electronic structure of lithium at lower pressure in the $I\bar{4}3d$ phase(see FIG. S1), which is also nonsymmorphic. However, the eight-fold degeneracies at the H point, predicted in the band structure of any crystal with $I\bar{4}3d$ symmetry \cite{Wieder_et_al:2016}, are far below the Fermi energy and topologically-trivial nearly-free electron-like bands dominate the electronic structure at the Fermi level. 

Our computed Fermi surface of the $Pbca$ phase at 80 GPa appears in FIG. \ref{fig:pbca}(c). The fourfold degenerate Dirac points form part of a nodal ring located at the Fermi energy in the $k_z$=0 plane of the Brillouin zone, enforced by the glide plane \{2$_{001}$ | $\nicefrac{1}{2}\ 0\ \nicefrac{1}{2}$\} (in Seitz notation). Our calculations predict that the nodal ring is well-isolated from the nearest bands over a broad energy range of 0.8 eV. We verify that the nodal ring is protected by a nonzero Berry phase (winding number = --1), and therefore it is topologically nontrivial (see SI for details). The Fermi velocities, computed with DFT-LDA, range from 2.8--6.6x$10^5$ m/s, comparable to measured values for other verified Dirac semimetals, such as Na$_3$Bi \cite{Liu_et_al:2014b} and Cd$_3$As$_2$ \cite{Neupane_et_al:2014}. Topologically nontrivial electronic bands will lead to unique surface states arising from the bulk topological features. Although such surface states would be ostensibly challenging to probe experimentally, our DFT calculations of the \textit{Pbca} (001) surface-projected band structure (FIG. \ref{fig:pbca}(d)) verify the existence of the expected drumhead surface state bands connecting the bulk Dirac points \cite{Burkov_et_al:2011,Rui_et_al:2018}.

\begin{figure}
\includegraphics[width=\linewidth]{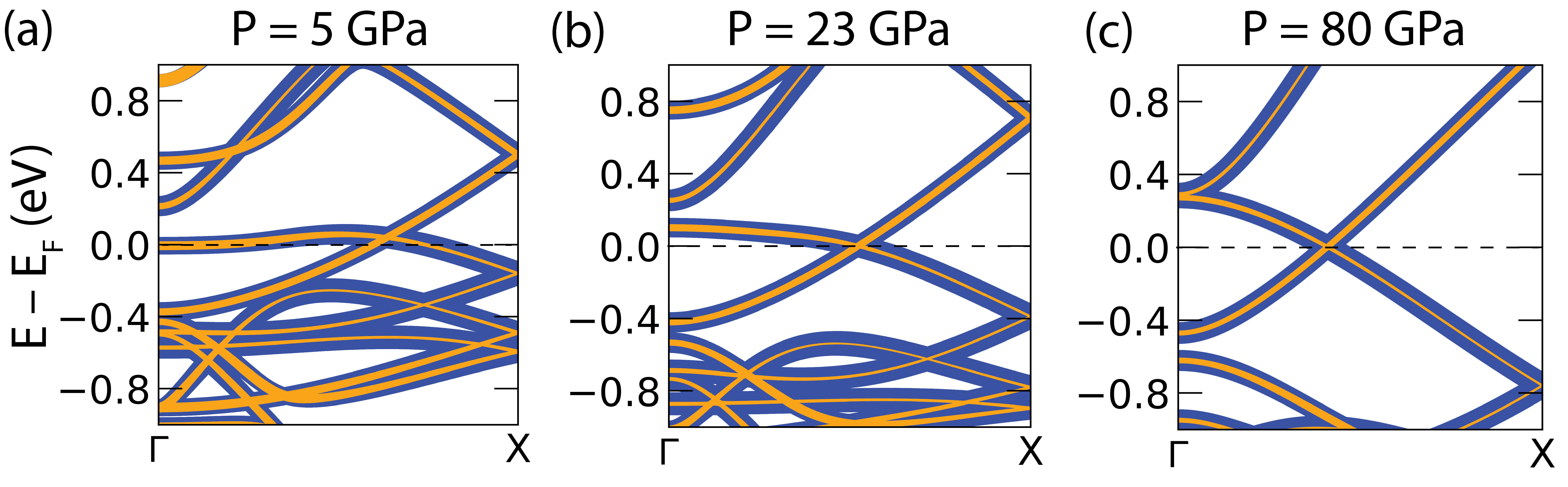}
\caption{\label{fig:pbcapressure} Band structures computed with DFT-LDA for bulk \textit{Pbca} phase shown in order of increasing pressure from (a) 5 GPa through to (c) predicted pressure where the space group is stable at 80 GPa. The relative orbital contribution to the bands is shown where s-like is in orange and p-like in blue. Increasing pressure shows increasing p character in the bands at the Fermi level, the s-like contribution decreases from \textasciitilde 30\% to 10\% comparing (a) and (c). Along with the broadening of bands this gives rise to a well-isolated Dirac crossing. A similar transition is seen along the $\Gamma$--Y direction.} 
\end{figure}

That high-pressure phases of solid Li possess topological band structures with pressure is notable, given its low atomic number. From a geometric perspective, we note that the \textit{Pbca} structure can be viewed as consisting of distorted honeycomb layers, in approximate analogy with graphene, possessing two distortions relative to the pristine honeycomb lattice: the first is a buckling distortion that results in neighboring Li atoms displacing in opposite directions to each other, perpendicular to the atomic plane; the second is a bond stretching distortion that results in each lithium atom having one shorter and two longer bonds with its nearest neighbors. 

To understand the origin of the Dirac crossings and topological nature of the \textit{Pbca} phase, we perform a computational experiment. Keeping the symmetry and Wyckoff positions fixed, we alter the volume of Li in the \textit{Pbca} phase (see SI for lattice parameters). We consider several lower pressures between 5 GPa and 80 GPa, the latter the pressure at which this phase is expected to be first preferred. In FIG. \ref{fig:pbcapressure} we see that along the $\Gamma$--X direction, there is a band crossing close to the Fermi energy. As the pressure increases, the initially flat band shows increasing p-like character, leading to broadening of the band and a change in curvature where the band is now higher in energy at the zone center ($\Gamma$) than further towards the zone edge, until the band crossing point. An increase in p-orbital character in the near-Fermi energy band structure of Li and other alkali metals under pressure has been noted before \cite{Boettger/Trickey:1985,Boettger/Albers:1989,Neaton_et_al:2001}. This change in band character broadens the energy range over which we have linearly dispersing bands, and pushes trivial bands further from the Fermi level ensuring the nodal ring is well isolated from other bands. Although we only show the $\Gamma$--X direction in Fig. \ref{fig:pbcapressure}, we compute the same effect for the Dirac node along the $\Gamma$--Y direction as well. The increasing p-character in the bands in combination with the underlying nonsymmorphic crystalline symmetry gives rise to isolated Dirac crossings along these high symmetry lines at the Fermi level.

\begin{figure}
\includegraphics[width=\linewidth]{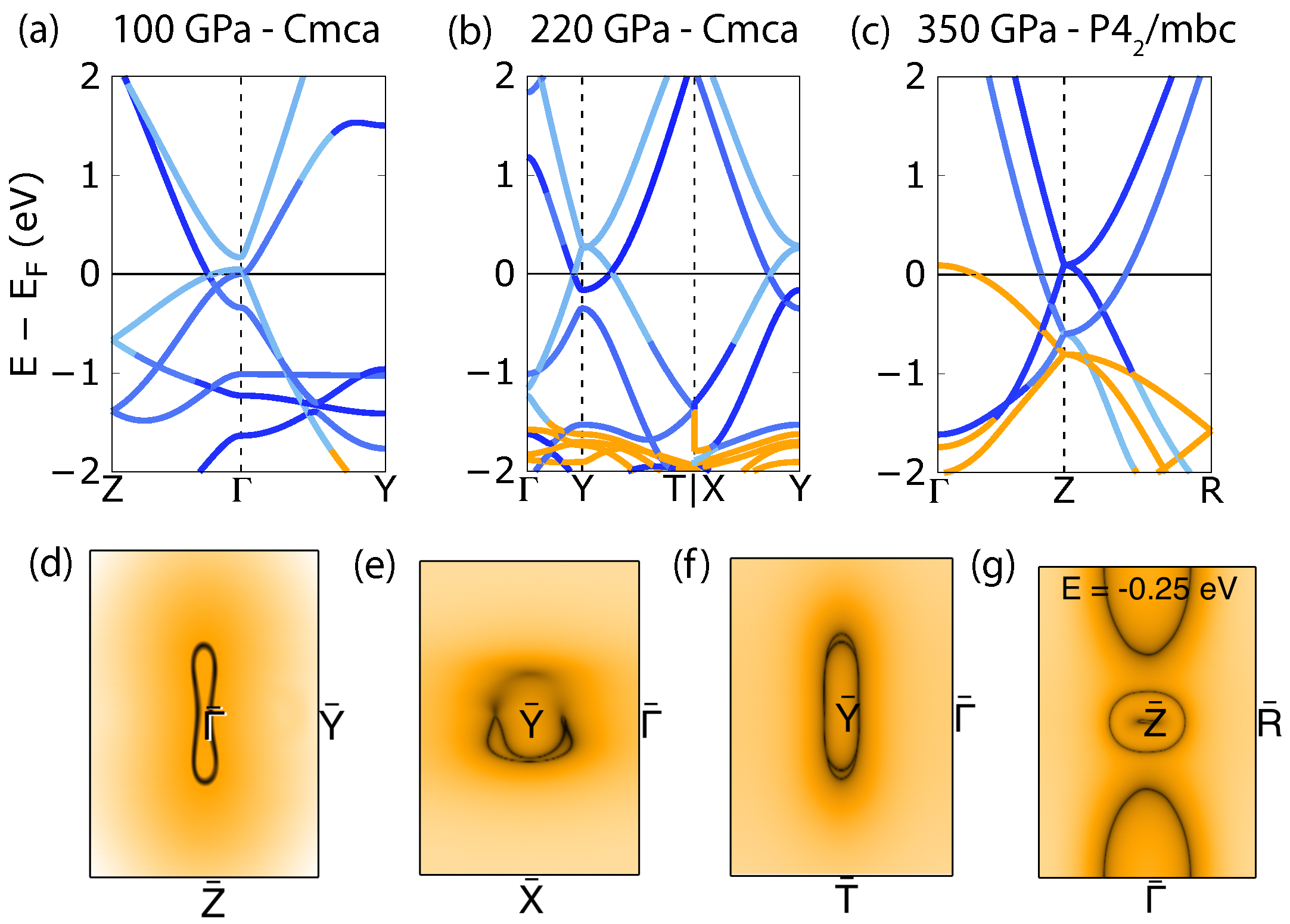}
\caption{\label{fig:cmca-p42mbc} Band structures for higher pressure bulk phases for (a) \textit{Cmca-24}; (b) \textit{Cmca-56}; and (c) \textit{P4$_2$/mbc}. The relative orbital contribution to the bands is shown where s-like is in orange and p-like in blue. The Dirac nodes are still present in the band structures, although as the pressure increases the s-like bands rise in energy relative to the Fermi level and the topological features are not as well-isolated from the trivial bands as in the \textit{Pbca} case. The corresponding calculated Fermi surfaces projected on a 2D plane in the Brillouin zone are shown in (d)-(g) (The full 3D Fermi surfaces are shown in the SI.) (d) \textit{Cmca-24}: in the k$_z$=0 plane, the nodal loop is largely derived from the bands close to the $\Gamma$ point and is much smaller in k-space than for the \textit{Pbca} nodal ring; (e) \textit{Cmca-56}: there are two nodal rings. The first nodal loop depicted in (e) is derived from the two crossings seen in the electronic band structure from $\Gamma$--Y and Y--X; the second nodal loop in (f) is spanned by the high symmetry directions from $\Gamma$--Y and Y--T seen at the Fermi energy in the electronic band structure.; (g) \textit{P4$_2$/mbc}: at an energy 250 meV below E$_F$ there are two nodal loops centered at Z formed of the p-like bands as seen in the linear crossings in the corresponding band structure plot (c) at the same energy.}
\end{figure}

We compute that the band structures of the next three predicted phases at higher pressures -- \textit{Cmca-24}, \textit{Cmca-56}, and \textit{P$4_2$/mbc} -- also exhibit Dirac-like bands close to the Fermi energy (FIG. \ref{fig:cmca-p42mbc}). While all three space groups -- \textit{Pbca, Cmca, P4$_2$/mbc} --  feature space groups with nonsymmorphic symmetries (glide planes) and inversion, the difference between \textit{Pbca} and \textit{Cmca} is the introduction of additional translation symmetry, and \textit{P4$_2$/mbc} includes a screw axis and reflection plane. Examining the computed Fermi surfaces of these structures at 100 GPa and 220 GPa, respectively, we note our DFT calculations predict that the \textit{Cmca-24} phase has a Fermi surface composed of a `pinched' nodal ring (viewed along [001] in FIG. \ref{fig:cmca-p42mbc}(d)) with two flat lobes adjacent, but disconnected, at 100 GPa (see FIG. S3). Interestingly, this pinched nodal ring evolves into two perpendicular nodal rings as Li transitions to \textit{Cmca-56} at 220 GPa; these are not isoenergetic so appear faded in the top half of FIG. \ref{fig:cmca-p42mbc}(e), and the second loop is spanned by the high symmetry directions from $\Gamma$--Y and Y--T as shown in FIG. \ref{fig:cmca-p42mbc}(f). This Lifshitz transition \cite{Lifshitz:1960,Volovik:2017}, or change in topology of the Fermi surface, can be seen in the constant energy cuts in FIG. \ref{fig:cmca-p42mbc} (d) -- (f) as it evolves from a single loop centered at $\Gamma$ to a double loop centered at the Y point. A Lifshitz transition was initially suggested to occur in metals under pressure \cite{Lifshitz:1960}; while it has been observed experimentally in heavier elements \cite{Andrianov:2014,Dipietro_et_al:2018,Sun_et_al:2017,Gaidukov_et_al:1979,Brandt_et_al:1985,Overcash_et_al:1981,Gaidukov_et_al:1977,Budko_et_al:1984,Watlington_et_al:1977,Schirber_et_al:1971,Volynskii_et_al:1984,Chu_et_al:1970}, we predict it to occur in lithium as well.

The nontrivial topological features for solid Li persist in the predicted high-pressure phases from 80--500 GPa, albeit with increasing contributions from trivial bands around the Fermi level as compared to \textit{Pbca}. At higher pressures of 350 GPa, the near-Fermi energy Li band structure reverts to having more and more s-like character (shown in orange in FIG. \ref{fig:cmca-p42mbc} (c)), and in fact s-like character dominates at the Fermi level in the band structure of the \textit{P4$_2$/mbc} phase. Computing the electronic structure of this phase at 350 GPa, we find that the nodal rings are still present in this phase, but are now approximately 0.25 eV below the Fermi level in our DFT calculations (see SI). This is expected since the p-like bands are broadened more significantly under pressure reverting the order of the s- and p-like bands \cite{Zittel_et_al:1985}; this has the effect of moving the Dirac nodes to lower energies relative to the Fermi level. As a growing number of states are present at the Fermi level, lithium has increasing metallic character at higher pressures compared to the \textit{Pbca} phase at 80 GPa. 

\begin{figure}
\includegraphics[width=\linewidth]{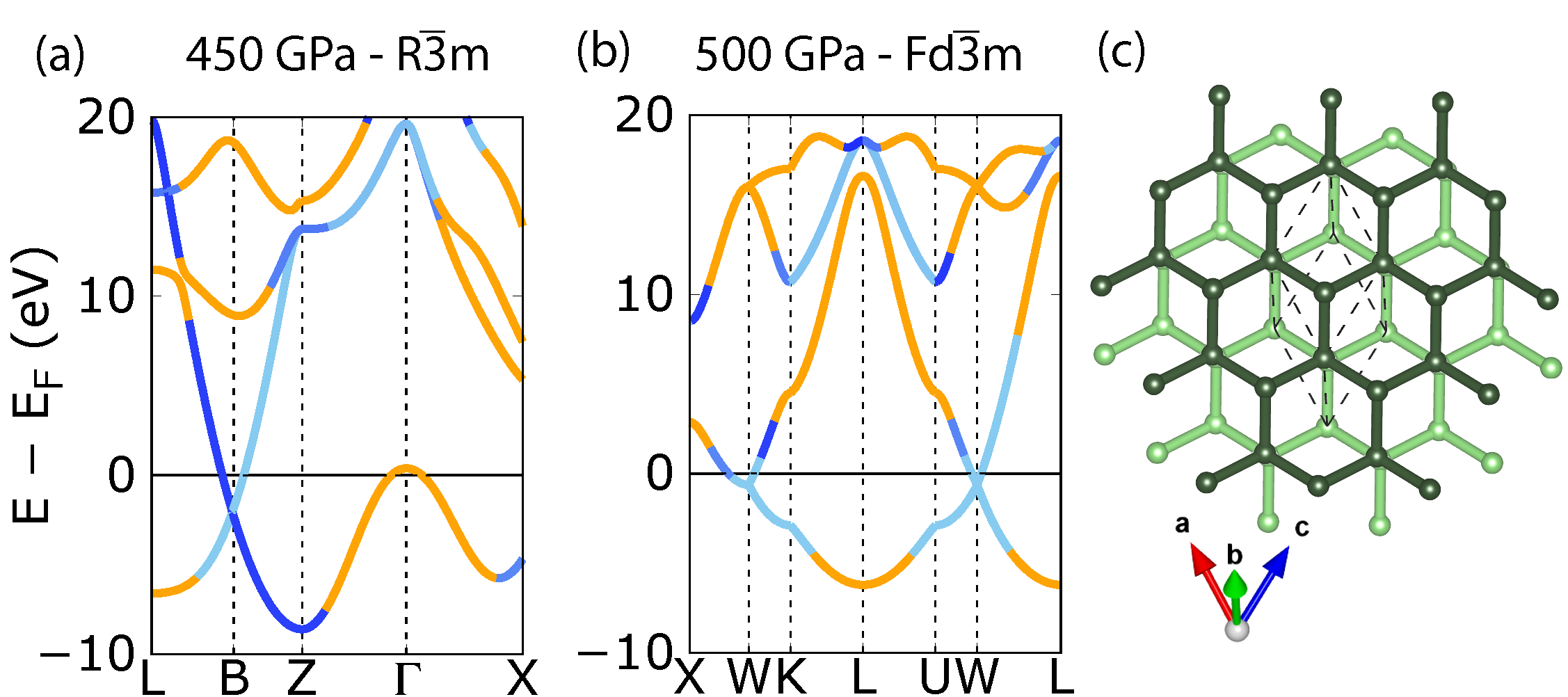}
\caption{\label{fig:fd3m} Band structures for higher pressure bulk phases for (a) \textit{R$\bar{3}$m}; and (b) \textit{Fd$\bar{3}$m} where the coloring follows the same scheme as Fig. \ref{fig:pbca}(b) and \ref{fig:cmca-p42mbc}. We note the graphene-like Dirac crossing at the W point 1 eV below E$_F$ in (b) and the bands have p-character. (c) \textit{Fd$\bar{3}$m} unit cell - coloring indicates different atomic layers which have the same buckling distortion perpendicular to the atomic plane in the hexagonal motifs similar to the \textit{Pbca} structure;  here the atoms are four-fold coordinated with their nearest neighbors and all bond lengths are 1.23 $\mbox{\AA}$.}
\end{figure}

In the final two phases of the predicted pressure phase diagram, the lithium atoms are four-fold coordinated and the electronic structure exhibits highly-dispersive bands and metallic character at the Fermi level (Fig. \ref{fig:fd3m}). The \textit{R$\bar{3}$m} phase is symmorphic and its electronic structure is dominated by nearly-free electron-like bands. But as lithium adopts the \textit{Fd$\bar{3}$m} structure, the similarity to graphene becomes more pronounced in both its geometry and electronic structure: its structure consists of offset, hexagonal layers with all bond lengths approximately 1.23 Å (both within each hexagonal layer and between layers) and has the same `buckling' motif, as described earlier for the \textit{Pbca} phase. We note that at the W point on the Brillouin zone edge, where the bands are dominated by p-like character, there is a Dirac point, similar to graphene's Dirac point at its K point (which also lies on the Brillouin zone edge); albeit, here in lithium, it is 1 eV below the Fermi energy whereas in graphene it is at the Fermi level. Lithium can thus be seen to form a 3D analogue to graphene at this extremely high pressure, where again the crystallographic symmetry and dominant p-character at the Fermi level facilitates the formation of the Dirac crossing.

We note that at low temperatures, topological properties are predicted appear at about 80 GPa and above, a pressure range recently probed for lithium \cite{Guillaume_et_al:2011}; pressures up to 400 GPa are experimentally achievable using diamond anvil cells \cite{Li_et_al:2018}, and 500 GPa appears to be within reach in the near future \cite{McMahon_et_al:2012,Dias/Silvera:2017,Geng:2017}. The closed loop in the Fermi surfaces, would lead to characteristic quantum oscillations in de Haas-van Alphen measurements for the \textit{Pbca}, \textit{Cmca}, and \textit{P4$_2$/mbc} structures (shown in the SI). Given the small lithium mass, nuclear motion associated with zero-point and finite temperature effects would be expected to alter the structural phase diagram \cite{Deemyad/Zhang:2018,Ackland_et_al:2017,Elatresh_et_al:2017}; such effects may shift the zero-temperature structural phase transition pressures predicted in \cite{Lv_et_al:2011,Pickard/Needs:2009} and used here. Future calculations, for example including anharmonic effects in calculations of structural energetics, as has recently been demonstrated for hydrogen \cite{Monserrat_et_al:2018}, would be desirable.  Additionally, appreciable electron-phonon interactions expected for lithium at high densities could potentially modify band dispersion, introducing satellites and kinks as has been reported in angle-resolved photoemission studies of graphene \cite{Park_et_al:2007,Bostwick_et_al:2007}. Further theoretical and experimental studies will be important to explore these issues in detail and their consequences for the structural and electronic phase behavior of lithium at these high densities.

In summary, as lithium is subjected to higher and higher pressures, it favors lower coordinated phases with nonsymmorphic symmetries which enforce band stickings at high symmetry points, thus increasing the likelihood for crossings along high symmetry directions.  The distorted hexagonal honeycomb structural motifs are reminiscent of graphene and are consistent with the nonsymmorphic symmetries that guarantee the presence of band stickings. Furthermore, the pressures at which these structures are favored not only reorder the bands and lead to dominant p-orbital character near the Fermi level, but also shifts many of the trivial electronic bands away from the Fermi energy as they become more significantly broadened. These two effects in combination, the dominant p-character and the phase transitions to structures with nonsymmorphic symmetries, result in well-isolated Dirac nodes at the Fermi energy in high pressure lithium. When lithium assumes the \textit{Cmca} symmetry, it undergoes a Lifshitz transition as seen by the change in Fermi surface topology between the two predicted structures. It then evolves to a more metallic \textit{P4$_2$/mbc} phase where the nodal line is slightly below the Fermi energy as the s- and p-like bands become reordered with pressure. At 500 GPa, lithium's structure consists of four-fold coordinated atoms in buckled hexagonal honeycomb layers, giving rise to a Dirac crossing 1 eV below the Fermi level as predicted by our DFT calculations. Using first-principles calculations, we show here that lithium's complex structural phase diagram also features topological electronic structure, suggesting similar features may be observed in other light elements under pressure. Indeed, more generally, pressure can be used to realize topological features in electronic structures in broad classes of materials.

 \matmethods{

 \subsection*{Computational Methods}

Density functional theory calculations are carried out with the Vienna \textit{ab initio} simulations package \cite{Kresse/Hafner:1993,Kresse/Furthmuller:1996,Kresse/Joubert:1999} (VASP) within the local density approximation (LDA). We use a plane-wave basis set and projector augmented-wave pseudopotentials which, for Li, treat both 1s and 2s electrons explicitly as valence. Our plane-wave energy cutoff is 1000 eV (to achieve energy convergence across all the high pressure structures) and a k-grid density of 0.01 {\AA}$^{-1}$ is used. We find that including spin-orbit coupling has a negligible effect on the structural and electronic properties of the different phases and therefore all calculations are performed without spin-orbit coupling. Z2Pack is used to calculate the Chern number around the nodal ring using the evolution of Wannier charge center positions along some periodic direction of a surface in the Brillouin zone \cite{Soluyanov/Vanderbilt:2011}, in this case a torus constructed around the nodal ring. To examine surface states, tight-binding models are constructed using the hopping parameters obtained from the Wannier functions calculated using Wannier90 \cite{Mostofi_et_al:2014}. As implemented in the open-source package WannierTools \cite{Wu_et_al:2017}, the Fermi surface is calculated via our tight-binding analysis and the surface density of states is calculated using an iterative Green's function technique \cite{Sancho_et_al:1985}.  
 }

\showmatmethods{} 

\acknow{This work was supported by the Theory FWP at the Lawrence Berkeley National Laboratory, which is funded by the U.S. Department of Energy, Office of Science, Basic Energy Sciences, Materials Sciences and Engineering Division under Contract No. DE-AC02-05CH11231. Work performed at the Molecular Foundry was also supported by the Office of Science, Office of Basic Energy Sciences, of the U.S. Department of Energy under the same contract number. This research used resources of the National Energy Research Scientific Computing Center, a DOE Office of Science User Facility supported by the Office of Science of the U.S. Department of Energy, again under the same contract number. S.M.G. acknowledges financial support by the Swiss National Science Foundation Early Postdoctoral Mobility Program.}

\showacknow{} 

\bibliography{steph}

\end{document}